\documentclass{elsart}
\usepackage{amsmath}
\usepackage{amssymb}
\usepackage{graphics}
\begin{document}

\begin{frontmatter}
\title{Eta Photoproduction as a Test of the\\
Extended Chiral Symmetry}
\author[IEM]{C. Fern\'andez-Ram\'{\i}rez\corauthref{cor}},
\corauth[cor]{Corresponding author. Fax: +34915855184}
\ead{cesar@nuc2.fis.ucm.es}
\author[IEM,UCM]{E. Moya de Guerra},
\author[UCM]{J.M. Ud\'{\i}as}
\address[IEM]{Instituto de Estructura de la Materia, CSIC, 
Serrano 123, E-28006 Madrid, Spain}
\address[UCM]{Departamento de F\'{\i}sica At\'omica, Molecular 
y Nuclear, Facultad de Ciencias F\'{\i}sicas, 
Universidad Complutense de Madrid, Avda. Complutense s/n, 
E-28040 Madrid, Spain}

\begin{abstract}
We analyze the $\gamma p \to \eta p$ process from threshold up to 1.2 GeV,
employing an effective Lagrangian approach
that allows for a mixing of eta couplings of pseudoscalar and pseudovector nature.
The mixing ratio of the couplings may serve as a quantitative estimation
of the $SU_L(3)\times SU_R(3)$ extended chiral symmetry  violation in this energy regime.
The data analyzed (differential cross sections and asymmetries) 
show a preference for the pseudoscalar coupling
 --- 91\% of pseudoscalar coupling component for the best fit.
We stress that a more conclusive answer to this question
requires a more complete electromagnetic 
multipole database than the presently available one.
\end{abstract}
\end{frontmatter}

Our knowledge on effective field theories (EFTs) based upon 
Quantum Chromodynamics (QCD) has been enlarged thanks 
to the development and  success of chiral perturbation
theory ($\chi$PT) \cite{CHPT}. 
The starting point for $\chi$PT is QCD reduced to massless 
quarks $u$ and $d$. Under these assumptions the chiral 
$SU_L(2)\times SU_R(2)$ symmetry holds and
is spontaneously broken into $SU_I(2)$ symmetry, emerging
the nucleon as the ground state of the EFT and the
pion as the Goldstone boson that mediates 
the strong interaction in the EFT.

Even if the quarks $u$ and $d$ masses are strictly nonzero, they are small 
and relatively close, so that  chiral symmetry is a well-established approximate
symmetry of the strong interaction.
Indeed, 
chiral symmetry requires a vanishing $\pi NN$ coupling
for vanishing pion momentum.
The simplest way to achieve this is
by means of a pseudovector coupling to the pion
 \cite{Burgess}.
The extension of $\chi$PT to strange particles has been investigated
in the last years 
\cite{CHPT2,Kaiser95,Kaiser97,Kolomeitsev,Inoue,Nieves,Doring06}
with the extended chiral symmetry (E$\chi$S) 
$SU_L(3)\times SU_R(3)$ whose validity is more debatable.
This symmetry is spontaneously broken into $SU_V(3)$ symmetry and
eight Goldstone bosons emerge: 
four non-strange pseudoscalar mesons 
($\pi^\pm$, $\pi^0$, and $\eta$) and
four strange pseudoscalar mesons ($K^\pm$, $K^0$, and $\bar{K}^0$).
One can appreciate that 
the E$\chi$S is not restored as a good symmetry at the same level as standard
 chiral symmetry looking at the differencies among the masses of the meson 
octect \cite{PDG2006}.
The question arises about how important this breakdown of E$\chi$S arises
 already at the level of EFT or whether any restoration effect might happen 
in reactions, beyond the mass differences of hadrons or mesons belonging to 
E$\chi$S multiplets.

This Letter is devoted to the $\gamma p \to \eta p$ process
and the nature of the coupling of the $\eta$ particle, as
a possible  indication of violation of the E$\chi$S.
The study of the eta photoproduction process is well motivated
from both theoretical and experimental points of view. This is due to the
relation of the eta to the E$\chi$S and the study of the electromagnetic 
properties of the nucleon excitations as well as to the
experimental programs on eta photoproduction
developed in several worldwide facilities over 
the last years \cite{experiments,Price,Krusche,Renard}.
We focus on the nature of the coupling of the eta 
to the nucleon and its excitations,
mainly the N(1535) which presents a large $\eta N$ branching ratio
and dominates the threshold behavior of the eta photoproduction reaction. 
Whereas the pion coupling is usually chosen pseudovector because of the
chiral symmetry, there is no compelling  
\textit{ab initio} reason to decide that the $\eta$ meson has to be either
pseudoscalar (PS) or pseudovector (PV). However,
if the E$\chi$S were exactly fulfilled, the eta would couple to
the nucleon and its excitations through purely PV couplings with no PS admixture,
as in the case of pions.
Hence, if
the PS-PV mixing ratio $\varepsilon$ of the coupling could be reliably derived
from experimental data, it might provide a quantitative
indication of  E$\chi$S breakdown in the nucleon mass energy regime.
With this aim,  we explore whether the $\eta NN$ and $\eta N N^*$ vertices are 
either of PS, PV, or PS-PV mixing nature by fitting an effective Lagrangian model to data.

One must be well aware that in meson-baryon to meson-baryon processes
PS and PV couplings are indistinguishable, 
because they provide the same amplitude \cite{Benmerrouche, Gridnev,Friar}.
Thus, meson photoproduction emerges as the ideal way to distinguish between both 
couplings. The contribution of the anomalous magnetic moment of the proton 
makes a difference in the amplitudes \cite{Benmerrouche} originating 
from PV or PS couplings. 
We study whether, within the presently available experimental information, it is 
feasible to extract the possible PS contribution to the amplitude and test 
the E$\chi$S in the meson sector.

E$\chi$S is at the basis of an extensive study of
the nucleon and its excitations by means of 
eta photoproduction \cite{Kaiser97,Doring06} and
meson scattering \cite{Inoue} using chiral unitary approaches.
Coupled-channel chiral unitary models have been succesful in describing certain resonances as dynamically generated states, for instance, N(1535) \cite{Kaiser95,Inoue,Nieves}, 
N(1520) \cite{Kolomeitsev} and N(1650) \cite{Nieves}.
However, the nature of the N(1535), N(1520) and N(1650) is debatable.
Standard quark models \cite{Capstick00}, for instance, provide the right quantum 
numbers and mass predictions \cite{Doring06} for these baryon states.
Despite of the success of these models, one must be aware of the fact
that meson-baryon scattering cannot be used to test E$\chi$S.
The possible PS-PV mixing cannot be unveiled by these
processes due to the fact that, on-shell, both the PV 
and the PS couplings yield the same amplitudes for meson-baryon scattering.
As already mentioned, to study the possible PS-PV mixing one has to explore
photoproduction reactions.
The eta photoproduction process has been described within the chiral unitary approach with different
results. In Ref. \cite{Kaiser97}, the authors succeeded in reproducing the total cross section within 
a computation which allows for terms in the Lagrangian that explicitly break the E$\chi$S, while
the computation of the total cross section in \cite{Doring06} needs 
further improvements to achieve results in as good agreement with the data as the ones presented in \cite{Kaiser97} and in this Letter. 

In this Letter we present a realistic model of the $\gamma p \to \eta p$ 
reaction based upon the effective Lagrangian approach (ELA),
that from the theoretical point of view is a very suitable 
and appealing method to study meson photoproduction 
and nucleon excitations.
Recently, we presented a model for pion photoproduction on free
nucleons based on ELA \cite{fernandez06}.
In this Letter, we extend this model to the
eta photoproduction process.
The model includes Born terms and $\omega$ and $\rho$ 
vector mesons as well as
nucleon resonances up to 1.8 GeV mass and up to spin-3/2,
covering the energy region from threshold up to 1.2 GeV of
photon energy in the laboratory frame.
The model is fully relativistic and displays 
gauge invariance and crossing symmetry among other relevant features
that will be pointed out throughout this Letter.

We choose the $\eta NN$ interaction to be 
\begin{equation}
{\mathcal L}_{\eta NN}^{\text{PS-PV}}= g_{\eta NN} \left[
\frac{\left( 1-\varepsilon \right)}{2M}
\bar{N}\gamma_\alpha \gamma_5  N 
\partial^\alpha \eta 
-i\varepsilon \bar{N}\gamma_5 N \eta
\right] ,
\end{equation}
where $M$ is the mass of the nucleon, 
$g_{\eta NN}$ is the $\eta NN$ coupling constant,
and $\varepsilon$ is the mixing parameter which runs from 0 to 1.
Both $g_{\eta NN}$ and $\varepsilon$ will be fitted to data
within their physical ranges. 
The pure PS ($\varepsilon=1$) and PV ($\varepsilon=0$) choices
for the $\eta NN$ coupling have been explored  in Ref. \cite{Benmerrouche}.
In this Letter we go further,
not only because we allow for a mixing of PS and PV couplings through
the parameter $\varepsilon$ but also because we apply
the mixing to both Born terms and nucleon resonance contributions.

The mixing idea has been previously used in studies of the effect of
meson-exchange currents in muon capture by $^3$He \cite{truhlik} 
and to pion scattering \cite{Gridnev}. The latter
obviously provides the same result for both couplings because
in the absence of the contribution from the
anomalous magnetic moment of the nucleon,
PS and PV couplings yield the same amplitudes \cite{Friar}.
 
The electromagnetic coupling to the nucleon ($\gamma NN$ Lagrangian)
is given by
\begin{equation}
\begin{split}
{\mathcal L}_{\gamma NN}=&
-e\hat{A}^\alpha \bar{N} \gamma_\alpha \frac{1}{2} 
\left( 1+\tau_3 \right) N \\
&-\frac{ie}{4M} \bar{N} \frac{1}{2} 
\left( \kappa^S+\kappa^V \tau_3 \right) \gamma_{\alpha \beta} N
F^{\alpha \beta} ,
\end{split}
\end{equation}
where $e$ is the absolute value of the electron charge, 
$F^{\alpha \beta}=\partial^\alpha \hat{A}^\beta
-\partial^\alpha \hat{A}^\beta$ is the electromagnetic field, 
$\hat{A}^\alpha$ is the photon field,
and $\kappa^S=\kappa^p+\kappa^n=-0.12$ and
$\kappa^V=\kappa^p - \kappa^n=3.70$
are respectively 
the isoscalar and the isovector anomalous magnetic moments 
of the nucleon.

Vector mesons are included through the standard Lagrangians 
($V= \omega , \rho$) \cite{fernandez06}
\begin{equation}
\begin{split}
{\mathcal L}_V = &-F_{V N N}\bar{N}\left[\gamma_\alpha 
-i\frac{K_V}{2M}\gamma_{\alpha \beta}\partial^\beta \right]
V^\alpha N \\
&+\frac{e G_{V \eta \gamma}}{m_\eta} 
\tilde{F}_{\mu \nu} \left( \partial^\mu \eta \right)
V^\nu ,
\end{split}
\end{equation}
with $\tilde{F}^{\mu \nu}=\frac{1}{2}
\epsilon^{\mu \nu \alpha \beta} F_{\alpha \beta}$.
From Particle Data Group \cite{PDG2006} we take
the following values: 
$m_\eta=547.3$ MeV, $m_{\rho^0}=768.5$ MeV,
$m_\omega=782.57$ MeV,
$G_{\rho^0 \eta \gamma} =1.06$ 
($\Gamma_{\rho^0 \eta \gamma}=0.062$ MeV), 
and $G_{\omega \eta \gamma} = 0.29$
($\Gamma_{\omega \eta \gamma}=0.005486$ MeV).  
From the analysis of the
nucleon electromagnetic form factors performed by Mergell, 
Mei\ss ner, and Drechsel \cite{MMD} we take the values 
$F_{\rho NN}=2.6$, 
$K_\rho =6.1 \pm 0.2$, 
$F_{\omega NN}=20.86 \pm 0.25$, 
and $K_\omega=-0.16 \pm 0.01$.

\begin{figure}
\begin{center}
\scalebox{0.46}[0.46]{\includegraphics{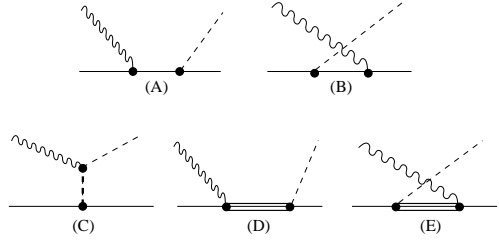}}
\end{center}
\caption{Feynman diagrams for Born terms: (A) s channel, 
(B) u channel; (C) vector-meson exchange terms; and
nucleonic excitations: (D) s channel and (E) u channel.} \label{fig:diag}
\end{figure}

Besides Born (diagrams (A) and (B) in Fig. \ref{fig:diag})
and vector meson exchange terms 
($\rho$ and $\omega$, diagram (C) in Fig. \ref{fig:diag}), 
the model includes six nucleon resonances:
N(1520), N(1535), N(1650), N(1700), N(1710), and N(1720)
--- diagrams (D) and (E) in Fig. \ref{fig:diag}.
Due to the isoscalar nature of the $\eta$ meson,
all the nucleon resonances involved are isospin-1/2.
An important virtue of the model employed here which overtakes 
former approaches lies in the treatment of the resonances.
We avoid well-known pathologies in the Lagrangians 
of the spin-3/2 resonances (such as N(1520)) 
of previous models, thanks to a modern approach 
due to Pascalutsa \cite{Pascalutsa}.
This coupling scheme has been applied succesfully 
to pion photoproduction from the nucleon and provides 
a better overall description of the behavior of the resonances 
\cite{fernandez06} than other spin-3/2 coupling schemes \cite{EMoya,Penner}.
Under this approach the (spin-3/2 resonance)-nucleon-eta and 
the (spin-3/2 resonance)-nucleon-photon
vertices have to fulfill the condition
$q_\alpha \mathcal{G}^{\alpha \dots}=0$ where $q$ is the four-momentum 
of the spin-3/2 particle, $\mathcal{G}^{\alpha \dots}$ stands for the vertex,
$\alpha$ the vertex index which couples 
to the spin-3/2 field, and the dots stand for other possible 
indices.
Within this prescription, the PS coupling to the eta yields
a zero amplitude contribution \cite{Pascalutsa}
and, thus, there is no 
PS-PV mixing for (spin-3/2 resonances)-nucleon-eta Lagrangians.
The D$_{13}$ Lagrangian, N(1520) and N(1700) resonances, can be written
\begin{equation}
\begin{split}
{\mathcal L}_{\text{D}_{13}}=&-\frac{H_{\eta NN^*}}{m_\eta M^*} \bar{N} 
\epsilon_{\mu \nu \lambda \beta} \gamma^\beta 
\left( \partial^\mu N^{*\nu} \right) 
\left( \partial^\lambda \eta  \right)  \\
&+ \frac{3e}{4M \left( M+M^* \right)} 
\bar{N} \left[ i \left( g_1^S+g_1^V \tau_3\right) 
\tilde{F}_{\mu \nu} \gamma_5 \right. \\
&+ \left. \left( g_2^S+g_2^V \tau_3 \right) 
F_{\mu \nu} \right] \partial^\mu N^{*\nu} + \text{H.c.} ,
\end{split} \label{eq:D13}
\end{equation}
where $M^*$ is the mass of the resonance, 
$\text{H.c.}$ stands for Hermitian conjugate,
$g_{1,2}^S$ and $g_{1,2}^V$ stand for the isoscalar and isovector
electromagnetic coupling constants respectively, 
and $H_{\eta NN^*}$ is the strong coupling constant.
The P$_{13}$ Lagrangian, N(1720) resonance, 
is obtained placing an overall $\gamma_5$ in (\ref{eq:D13}).

For S$_{11}$ resonances, N(1535) and N(1650), 
we build the PS-PV Lagrangian
\begin{equation}
\begin{split}
{\mathcal L}_{\text{S}_{11}}^{\text{PS-PV}}=&
\varepsilon iG_{\eta NN^*} \bar{N} N^* \eta
-\left( 1- \varepsilon \right) 
\frac{H_{\eta NN^*}}{m_\eta}\bar{N}\gamma_\alpha 
N^* \partial^\alpha \eta \\
&-\frac{ie}{4M}\bar{N}
\gamma_{\alpha \beta}\gamma_5\left(g_S+g_V\tau_3 \right) 
N^*F^{\alpha \beta} + \text{H.c.}
\end{split} \label{eq:S11}
\end{equation}
Despite the fact that we name two strong coupling constants 
$H_{\eta NN^*}$ and $G_{\eta NN^*}$, they only represent one parameter 
in the model because they are both related to the same 
experimental quantity, the partial decay width of the resonance 
into the $\eta N$ state, $\Gamma_\eta$.
The P$_{11}$ Lagrangian, N(1710) resonance is obtained 
placing an overall $\gamma_5$ in (\ref{eq:S11}).

Dressing of the resonances is considered by means of a 
phenomenological width which takes into account 
decays into one pion, one eta and two pions \cite{fernandez06}.
The phenomenological width employed is an improvement of those used by
Manley et al. \cite{Manley} (inspired by Blatt and Weisskopf factors \cite{Blatt})
and Garcilazo and Moya de Guerra \cite{EMoya}.
This width is energy dependent and is built so that it fulfills crossing symmetry and contributes to both direct and crossed channels of the resonances.
It also accounts for the right angular barrier of the resonance at threshold.
Consistency requires to incorporate the energy dependence of the
width in the strong coupling constants.
In order to regularize the high energy behavior of the model we include
a crossing symmetric and gauge invariant form factor for Born and
vector meson exchange terms, 
which contains the only free parameter in the model (the cutoff $\Lambda$)
together with
the mixing parameter $\varepsilon$.
The form factor for Born terms is \cite{Davidson01-1}
\begin{equation}
\begin{split}
\hat{F}_B(s,u,t)=& F(s)+F(u) +G(t)-F(s)F(u) \\
-& F(s)G(t)-F(u)G(t)+F(s)F(u)G(t) ,
\end{split}
\end{equation}
where,
\begin{eqnarray}
F(l)&=& \left[1+ \left( l-M^2 \right)^2/\Lambda^4 \right]^{-1} \: , 
\: l=s,u ;\\
G(t)&=& \left[1+ \left( t-m_\eta^2 \right)^2/\Lambda^4 \right]^{-1} ;
\end{eqnarray}
and for vector mesons we adopt $\hat{F}_V(t) = G(t)$ with 
the change $m_\eta \to m_V$. 

\begin{table}
\caption{Comparison among the PS, PV, and PS-PV prescriptions.
The subindex '$3$' stands for the fits where the constrain 
$g^2_{\eta NN}/4\pi=1.7$ has been imposed and the subindex '$0$' stands for
the fits where $g_{\eta NN}$ has been fitted to data. 
$dof$ stands for degrees of freedom and $\varepsilon$ for the mixing ratio.} 
\label{tab:varepsilon}
\begin{tabular}{lcccc}
\hline
         &$\varepsilon$ & $g^2_{\eta NN}/4\pi$ & $dof$ & $\chi^2/dof$ \\
\hline

PS$_3$       & 1        & 1.7                  & 633   & 15.37  \\  
PV$_3$       & 0        & 1.7                  & 633   & 18.19  \\
PS-PV$_3$& 0.99   & 1.7                 & 632   & 15.36  \\
PS$_0$       & 1        & 1.08               & 632   & 13.91  \\
PV$_0$       & 0         & 0.054             & 632   & 15.95  \\
PS-PV$_0$& 0.91   & 0.52               & 631   & 13.80  \\
\hline
\end{tabular}
\end{table}

In order to assess the parameters of the model we minimize the
function
\begin{equation}
\chi^2 = \sum_{j=1}^{n} \left[\frac{\mathcal{O}^{\text{experiment}}_j
-\mathcal{O}^{\text{model}}_j\left(\varepsilon, \Lambda, 
g_{\eta NN}, \dots \right)}
{\Delta \mathcal{O}^{\text{experiment}}_j} \right]^2 ,
\end{equation}
where $\mathcal{O}$ stands for the observables
--- namely differential cross sections and asymmetries 
(recoil nucleon polarization, polarized target, and polarized beam).
We use all
the available experimental database up to 1.2 GeV photon energy, 
a total amount of $n=665$ data points \cite{SAID}.
To perform the minimization we have used a 
genetic algorithm combined with the \texttt{E04FCF} routine
(gradient-based routine) from NAG libraries \cite{NAG}.
For each prescription of  the model we have allowed the parameters
(masses, widths, and electromagnetic coupling constants) 
to vary within the ranges given by the PDG independently.
Details on the optimization procedure applied can be found in 
\cite{thesis}.

\begin{figure}
\begin{center}
\rotatebox{-90}{\scalebox{0.32}[0.32]{\includegraphics{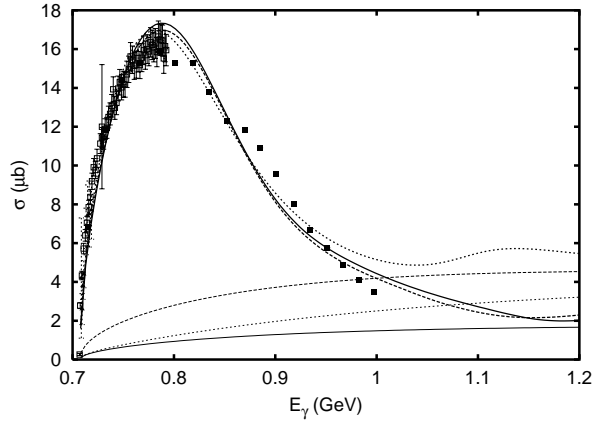}}}
\end{center}
\caption{Total cross section for the $\gamma p \to \eta p$ 
reaction. 
Thick lines: Full calculations;
Thin lines: B$+$VM contribution.
Solid, PS-PV$_0$ fit; 
Dashed, PS$_0$ fit; 
Dotted, PV$_0$ fit.
Data are from ELSA (solid triangles) \cite{Price},
MAMI (open squares) \cite{Krusche}, and
GRAAL (solid squares) \cite{Renard}.} 
\label{fig:xsec}
\end{figure}

\begin{figure}
\begin{center}
\rotatebox{-90}{\scalebox{0.32}[0.32]{\includegraphics{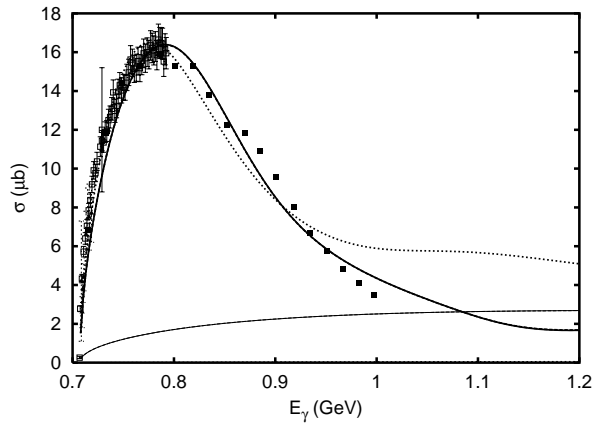}}}
\end{center}
\caption{Total cross section for the $\gamma p \to \eta p$ 
reaction with $g^2_{\eta NN}/4\pi=1.7$. 
Thick lines: Full calculations;
Thin lines: B$+$VM contribution.
Solid, PS-PV$_3$ fit; 
Dashed, PS$_3$ fit; 
Dotted, PV$_3$ fit.
Data as in Fig. \ref{fig:xsec}.
PS$_3$ and PS-PV$_3$ curves overlap.
The B$+$VM contribution to PV$_3$ curve is almost negligible.}
\label{fig:xsec3}
\end{figure}

In order to explore the reliability of the E$\chi$S
we have performed six fits using three different
prescriptions: PS ($\varepsilon=1$), PV ($\varepsilon=0$), and
PS-PV where $\varepsilon$ has been fitted to data.
Exact E$\chi$S predicts $g^2_{\eta NN}/4\pi=1.7$ \cite{Benmerrouche} so 
for each kind of fit we have
performed fits with and without this constrain.
The fits where this condition is imposed are named 
PS$_3$, PV$_3$, and PS-PV$_3$
and the fits where we let $g_{\eta NN}$ run freely within a sensible range
are named PS$_0$, PV$_0$, and PS-PV$_0$.
In Table \ref{tab:varepsilon} we provide a summary of our results.
We will report more extensive results obtained with our eta photoproduction model 
in a forthcoming publication \cite{eta}.
In advance, we provide in Figs. \ref{fig:xsec} and  \ref{fig:xsec3}
the results for the total cross section.

In the energy region near threshold,
the contribution of Born terms and vector mesons (B$+$VM in what follows) 
to the cross section
is small compared to the N(1535) resonance contribution, but as energy
increases the importance of B$+$VM contribution to the cross section
increases and eventually becomes the most important contribution.
Therefore, the relevance of the mixing parameter $\varepsilon$ stands out
mainly through the $\eta$-N(1535) vertex in the low-energy 
region and through the $\eta$N vertex in the high-energy region.
We find that the B$+$VM background to the
cross section is important -- at variance with Ref.  \cite{ETAMAID} --
and highly dependent on the coupling prescription.
In our model we do not account for 
final state interactions which might be important \cite{fernandez06}
in a reliable calculation/fit of the electromagnetic multipoles.

The PS and the PV prescriptions provide equally good agreement in the
near-threshold region, but as the energy increases and the Born terms
become more important, the PV prescription deviates from data
(see Figs. \ref{fig:xsec} and \ref{fig:xsec3}).
This fact together with the worse $\chi^2/dof$ supports PS coupling
as a better option for the phenomenological Lagrangian. Hence,
the E$\chi$S seems not to be an adequate underlying symmetry, at least 
for effective Lagrangians.
Differences between PS and PV coupling
are visible in the low energy 
region, not so much in the total cross section as in other observables that
are used to fit the parameters of the model, such as 
single polarization asymmetries and differential cross sections.

High-lying resonances contribute more 
in the high-energy region than in the low-energy one, but 
 they are not the main source of differences between results with PS and
 PV coupling for the total cross section. These resonances determine the shape 
 of the differential cross section and the asymmetries. Their
 contribution to the total cross section in the high energy region, though sizeable, is small 
 compared to the one of the tail of the N(1535) resonance plus Born terms 
 and vector mesons contributions.

The assumption of one single mixing parameter $\varepsilon$ is not the most 
general choice. In order to test this assumption we have repeated the fits to data with
two $\varepsilon$'s, one related to Born terms and another to the resonances.
We have performed the fits varying these parameters independently and 
no improvement in the $\chi^2/dof$ was found. The PS component remains dominant 
at the level of 90\% in both the resonances and the Born terms.

In summary, we have shown that eta photoproduction  is sensitive 
to the nature of the coupling and, thus, it is able to 
disentangle PS or PV couplings.
More high-quality experimental data, to
allow electromagnetic multipole separation, are 
needed in order to constrain the parameters of the model.
This will also allow to study final state interaction effects,
which becomes mandatory in order to attain further progress in this topic.
Meanwhile, the presented result for the dominance of the PS coupling
should be taken with caution.
Further research on the nature of the eta-nucleon-baryon coupling becomes mandatory.
Kaon photoproduction is  another interesting process where the
E$\chi$S can be tested following the lines presented in this Letter.

\begin{ack}
This work has been supported in part under contracts
FIS2005-00640 and FPA2006-07393
of Ministerio de Educaci\'on y Ciencia (Spain) and by UCM and Comunidad de Madrid
under project number 910059 (Grupo de F\'{\i}sica Nuclear).
Part of the computations of this work were carried out at the 
``Cluster de C\'alculo de Alta Capacidad para T\'ecnicas F\'{\i}sicas''
partly funded by EU Commission under program FEDER and by 
Universidad Complutense de Madrid (Spain).
\end{ack}


\begin{thebibliography}{99}
\bibitem{CHPT} V. Bernard, N. Kaiser, U.-G. Mei\ss ner, 
Nucl. Phys. B 383 (1992) 442-496; \\
H. Leutwyler, Ann. Phys. (N.Y.) 235 (1994) 165-203.
\bibitem{Burgess} C.P. Burgess, Phys. Rep. 330 (2000) 193-261.
\bibitem{CHPT2} V. Bernard, N. Kaiser, U.-G. Mei\ss ner,
Int. J. Mod. Phys. E 4 (1995) 193-344; \\
A.W. Thomas, W. Weise, 
\textit{The Structure of the Nucleon} (Wiley-VCH, Berlin, 2001).
\bibitem{Kaiser95} N. Kaiser, P.B. Siegel, W. Weise,
Phys. Lett. B 362 (1995) 23-28.
\bibitem{Kaiser97} N. Kaiser, T. Waas, W. Weise,
Nucl. Phys. A 612 (1997) 297-320.
\bibitem{Kolomeitsev} E.E. Kolomeitsev, M.F.M. Lutz,
Phys. Lett. B 585 (2004) 243-252; \\
S. Sarkar, E. Oset, M.J. Vicente Vacas,
Nucl. Phys. A 750 (2005) 294-323.
\bibitem{Inoue} T. Inoue, E. Oset, M.J. Vicente Vacas, Phys. Rev. C 65 (2002) 035204;
\bibitem{Nieves} J. Nieves, E. Ruiz Arriola, Phys. Rev. D 64 (2001) 116008.
\bibitem{Doring06} M. D\"oring, E. Oset, D. Strottman,
Phys. Rev. C 73 (2006) 045209.
\bibitem{PDG2006} W.-M. Yao \textit{et al.}, J. Phys. G 33 (2006) 1-1232.
\bibitem{experiments}
S.A. Dytman \textit{et al.}, Phys. Rev. C 51 (1995) 2710-2715; \\
A. Bock     \textit{et al.}, Phys. Rev. Lett. 81 (1998) 534-537; \\
J. Ajaka    \textit{et al.}, Phys. Rev. Lett. 81 (1998) 1797-1800; \\
M. Dugger   \textit{et al.}, Phys. Rev. Lett. 89 (2002) 222002; \\
B. Krusche, S. Schadmand, Prog. Part. Nucl. Phys. 51 (2003) 399-485; \\
V. Crede \textit{et al.}, Phys. Rev. Lett. 94 (2005) 012004.
\bibitem{Price} J.W. Price \textit{et al.}, Phys. Rev. C 51 (1995) 2283-2287.
\bibitem{Krusche} B. Krusche \textit{et al.},
Phys. Rev. Lett. 74 (1995) 3736-3739.
\bibitem{Renard} F. Renard \textit{et al.}, Phys. Lett. B 528 (2002) 215-220.
\bibitem{Benmerrouche} M. Benmerrouche, N.C. Mukhopadhyay, 
Phys. Rev. Lett. 67 (1991) 1070-1073; \\
M. Benmerrouche, N.C. Mukhopadhyay, J.F. Zhang,
Phys. Rev. D 51 (1995) 3237-3266.
\bibitem{Gridnev} A.B. Gridnev, N.G. Kozlenko,
Eur. Phys. J. A 4 (1999) 187-194.
\bibitem{Friar} J.L. Friar, B.F. Gibson, Phys. Rev. C 15 (1977) 1779-1782; \\ 
A.M. Bernstein, B.R. Holstein, Comments Nucl. Part. Phys. 20 (1991) 197-220.
\bibitem{Capstick00} S. Capstick and W. Roberts, 
Prog. Part. Nucl. Phys. 45 (2000) S241-S331.
\bibitem{fernandez06} C. Fern\'andez-Ram\'{\i}rez, 
E. Moya de Guerra, J.M. Ud\'{\i}as, 
Ann. Phys. (N.Y.) 321 (2006) 1408-1456; \\
C. Fern\'andez-Ram\'{\i}rez, E. Moya de Guerra, J.M. Ud\'{\i}as,
Phys. Rev. C 73 (2006) 042201(R); \\
C. Fern\'andez-Ram\'{\i}rez, E. Moya de Guerra, J.M. Ud\'{\i}as,
Eur. Phys. J. A 31 (2007) 572-574.
\bibitem{truhlik} J.G. Congleton, E. Truhlik, Phys. Rev. C 53 (1996) 956-976.
\bibitem{MMD} P. Mergell, U.-G. Mei\ss ner, D. Drechsel, 
Nucl. Phys. A 596 (1996) 367-396.
\bibitem{Pascalutsa} V. Pascalutsa, 
Phys. Rev. D 58 (1998) 096002.
\bibitem{EMoya} H. Garcilazo, E. Moya de Guerra, 
Nucl. Phys. A 562 (1993) 521-568.
\bibitem{Penner} G. Penner, U. Mosel, Phys. Rev. C 66 (2002) 055212.
\bibitem{Manley} D.M. Manley, R.A. Arndt, Y. Goradia, V.L. Teplitz, Phys. Rev. D 30 (1984) 904-936; \\
D.M. Manley, E.M. Saleski, Phys. Rev. D 45 (1992) 4002-4033.
\bibitem{Blatt} J. Blatt and V. F. Weisskopf, \textit{Theoretical Nuclear Physics} (Dover,
New York, 1991).
\bibitem{Davidson01-1} R.M. Davidson, R. Workman,
Phys. Rev. C 63 (2001) 025210; \\
R.M. Davidson, R. Workman, Phys. Rev. C 63 (2001) 058201.
\bibitem{SAID} R.A. Arndt, W.J. Briscoe, I.I. Strakovsky, 
R.L. Workman, Phys. Rev. C 66 (2002) 055213,
SAID database, http://gwdac.phys.gwu.edu
\bibitem{NAG} Numerical Algorithms Group Ltd.,
Wilkinson House, Jordan Hill Road, Oxford OX2-8DR, UK,
http://www.nag.co.uk
\bibitem{thesis} C. Fern\'andez-Ram\'{\i}rez, 
\textit{Electromagnetic production 
of light mesons}, PhD dissertation,
Universidad Complutense de Madrid (2006),
http://nuclear.fis.ucm.es/research/thesis/cesar\_tesis.pdf
\bibitem{eta} C. Fern\'andez-Ram\'{\i}rez, 
E. Moya de Guerra, J.M. Ud\'{\i}as, in preparation.
\bibitem{ETAMAID} L. Tiator, C. Bennhold, S.S. Kamalov, 
Nucl. Phys. A  580 (1994) 455-474;
W.-T. Ching, S.N. Yang, L. Tiator, D. Drechsel,
Nucl. Phys. A 700 (2002) 429-453.
\end{thebibliography}
\end{document}